\documentclass[sigconf, nonacm, pdfa]{acmart}

\newcommand\vldbdoi{ArXiv Version}
\newcommand\vldbpages{}
\newcommand\vldbvolume{18}
\newcommand\vldbissue{11}
\newcommand\vldbyear{2025}
\newcommand\vldbauthors{\authors}
\newcommand\vldbtitle{\shorttitle} 
\newcommand\vldbavailabilityurl{}

\usepackage{multirow}
\usepackage{booktabs} %
\usepackage{rotating}%

\usepackage{tikz}
\usepackage{array}
\usepackage{colortbl}

\begin{document}
	
	\title{TabulaX: Leveraging Large Language Models for Multi-Class Table Transformations}
	
	\author{Arash Dargahi Nobari}
	\email{dargahi@ualberta.ca}
	\affiliation{%
		\institution{University of Alberta}
		\city{Edmonton}
		\state{AB}
		\country{Canada}
	}
	
	\author{Davood Rafiei}
	\email{drafiei@ualberta.ca}
	\affiliation{%
		\institution{University of Alberta}
		\city{Edmonton}
		\state{AB}
		\country{Canada}
	}

	\renewcommand{\shortauthors}{Dargahi Nobari et al.}
	
	\begin{abstract}
		The integration of tabular data from diverse sources is often hindered by inconsistencies in formatting and representation, posing significant challenges for data analysts and personal digital assistants. Existing methods for automating tabular data transformations are limited in scope, often focusing on specific types of transformations or lacking interpretability. In this paper, we introduce TabulaX, a novel framework that leverages Large Language Models (LLMs) for multi-class column-level tabular transformations. TabulaX first classifies input columns into four transformation types---string-based, numerical, algorithmic, and general---and then applies tailored methods to generate human-interpretable transformation functions, such as numeric formulas or programming code. This approach enhances transparency and allows users to understand and modify the mappings. Through extensive experiments on real-world datasets from various domains, we demonstrate that TabulaX outperforms existing state-of-the-art approaches in terms of accuracy, supports a broader class of transformations, and generates interpretable transformations that can be efficiently applied.
	\end{abstract}


	\maketitle

	\pagestyle{plain}
	\begingroup\small\noindent\raggedright\textbf{PVLDB Reference Format:}\\
	\vldbauthors. \vldbtitle. PVLDB, \vldbvolume(\vldbissue): \vldbpages, \vldbyear.\\
%	\href{https://doi.org/\vldbdoi}{doi:\vldbdoi}
	\endgroup
	\begingroup
	\renewcommand\thefootnote{}\footnote{\noindent
		This work is licensed under the Creative Commons BY-NC-ND 4.0 International License. Visit \url{https://creativecommons.org/licenses/by-nc-nd/4.0/} to view a copy of this license. For any use beyond those covered by this license, obtain permission by emailing \href{mailto:info@vldb.org}{info@vldb.org}. Copyright is held by the owner/author(s). Publication rights licensed to the VLDB Endowment. \\
		\raggedright Proceedings of the VLDB Endowment, Vol. \vldbvolume, No. \vldbissue\ %
		ISSN 2150-8097. \\
		\href{https://doi.org/\vldbdoi}{doi:\vldbdoi} \\
	}\addtocounter{footnote}{-1}\endgroup
	
	\ifdefempty{\vldbavailabilityurl}{}{
		\vspace{.3cm}
		\begingroup\small\noindent\raggedright\textbf{PVLDB Artifact Availability:}\\
		The source code, data, and/or other artifacts have been made available at \url{\vldbavailabilityurl}.
		\endgroup
	}
	
	\vspace{40pt}
	
	\section{Introduction}
	The rapid growth of publicly available data and increased reliance on third-party sources have underscored the need for efficient methods to combine and transform data from diverse sources. This is crucial for a wide range of users, including data analysts, scientists, and personal digital assistants. However, significant challenges arise due to variations in formatting and presentation across these sources. Even within a single organization, data such as names, addresses, and financial records may exist in multiple inconsistent formats. Target tables may be stored in formats such as spreadsheets, relational databases, web tables, and comma-separated files---common across organizations, the web, and public government datasets~\cite{miller2018making,dtt}.

	There has been a wide array of studies on automating the transformation of tabular data and ensuring consistency across diverse data sources~\cite{dtt,gxjoin,TDE,DataXFormer}, yet significant challenges remain. Table~\ref{tab:topModelsOnKBWT} reports the performance, in terms of F1-score, of four state-of-the-art (SOTA) methods on KBWT~\cite{DataXFormer}, a dataset comprising tables extracted from a knowledge base (\S \ref{sec:datasets}). Clearly, all SOTA models struggle and only GXJoin generates explainable transformations, while the others produce either an output string or a match decision, both of which lack interpretability. 
	Figure~\ref{fig:etables} illustrates examples of source and target tables.
	In the first table pair, where individuals are identified by name in the source and username in the target, all SOTA models perform well. However, for the second pair, involving weight conversions between pounds and kilograms, and the third pair, linking company names with their CEOs, the models fail to produce meaningful results. While GPT-4o often produces correct outputs, reflected in its relatively high F1-Score in Table~\ref{tab:topModelsOnKBWT}, it does not yield transformations that are scalable to large datasets.

	\begin{table}[tbp]
		\centering
		\caption{Performance of SOTA models on KBWT dataset}
		\setlength\extrarowheight{2pt}
		\begin{tabular}{|l||c|c|c|c|}
			\toprule
			& GPT-4o & DTT~\cite{dtt} & GXJoin~\cite{gxjoin} & AFJ~\cite{auto-fuzzy-join}\\ \midrule
			F1-Score & 0.510 & 0.254 & 0.083 & 0.093\\
			Explainable & N & N & Y & N \\
			\bottomrule
		\end{tabular}%
		
		\label{tab:topModelsOnKBWT}
	\end{table}
	
	\begin{figure}[tbp]
		\centering
		\begin{tikzpicture}
\small
\setlength\extrarowheight{3pt}

	\definecolor{focuscolor}{HTML}{88d6e2}
	\definecolor{graycol}{rgb}{0.9,0.9,0.9}
	
	\def\targetX{3.0}
	\def\targetY{1.0}
	\def\OffsetY{4}
	\def\Y2{-\targetY-.6}
	\def\initX{-1.25}
	\def\SourceColWidth{2.75cm}
	\def\TargetColWidth{2.15cm}
	
	\def\otherColumnsHeader{\textbf{...}}

	\def\otherColumnsNum{1}
	\newcommand{\othercolumns}{
		\multicolumn{1}{>{\columncolor{graycol}}c|}{\textcolor{gray}{--}}
	}

	\newcommand{\PreserveBackslash}[1]{\let\temp=\\#1\let\\=\temp}
	\newcolumntype{C}[1]{>{\PreserveBackslash\centering}p{#1}}
	\newcolumntype{R}[1]{>{\PreserveBackslash\raggedleft}p{#1}}
	\newcolumntype{L}[1]{>{\PreserveBackslash\raggedright}p{#1}}

	\node at (\initX, 2.5) {\normalsize{\textbf{Source Table}}};
	\node at (\targetX, 2.5) {\normalsize{\textbf{Target Table}}};
	
	\node at (-3.6, \targetY) {\textbf{(1)}};
	\node at (-3.6, \Y2) {\textbf{(2)}};
	\node at (-3.6, -\targetY * \OffsetY) {\textbf{(3)}};

	\node (S1) at (\initX, \targetY) {
		\begin{tabular}{|C{\SourceColWidth}|c|c|c|}
			\hline
			\cellcolor{focuscolor}\textbf{Full Name} & \multicolumn{\otherColumnsNum}{>{\columncolor{graycol}}c|}{\otherColumnsHeader} \\
			\hline
			Nadia Ralph Allen & \othercolumns \\
			Sean Morse & \othercolumns \\
			\footnotesize{Dena Christopher Griffith} & \othercolumns \\
			Brandy Constable & \othercolumns \\
			\hline
		\end{tabular}
	};
	
	\node (S2) at (\initX, \Y2) {
		\begin{tabular}{|C{\SourceColWidth}|c|c|c|}
			\hline
			\cellcolor{focuscolor}\textbf{Weight in Pounds} & \multicolumn{\otherColumnsNum}{>{\columncolor{graycol}}c|}{\otherColumnsHeader} \\
			\hline
			2 & \othercolumns \\
			51.5& \othercolumns \\
			73 & \othercolumns \\
			\hline
		\end{tabular}
	};

	\node (S3) at (\initX, -\targetY * \OffsetY) {
		\begin{tabular}{|C{\SourceColWidth}|c|c|c|}
			\hline
			\cellcolor{focuscolor}\textbf{Company} & \multicolumn{\otherColumnsNum}{>{\columncolor{graycol}}c|}{\otherColumnsHeader} \\
			\hline
			Microsoft & \othercolumns \\
			PepsiCo & \othercolumns \\
			Toyota & \othercolumns \\
			\hline
		\end{tabular}
	};
	
	\node (T1) at (\targetX, \targetY) {
		\begin{tabular}{|C{\TargetColWidth}|c|c|c|}
			\hline
			\cellcolor{focuscolor}\textbf{Username} & \multicolumn{\otherColumnsNum}{>{\columncolor{graycol}}c|}{\otherColumnsHeader} \\
			\hline
			n.r.allen & \othercolumns \\
			s.morse & \othercolumns \\
			d.c.griffith & \othercolumns \\
			b.constable & \othercolumns \\
			\hline
		\end{tabular}
	};
	
	\node (T2) at (\targetX, \Y2) {
		\begin{tabular}{|C{\TargetColWidth}|c|c|c|}
			\hline
			\cellcolor{focuscolor}\textbf{Weight in Kg} & \multicolumn{\otherColumnsNum}{>{\columncolor{graycol}}c|}{\otherColumnsHeader} \\
			\hline
			0.9 & \othercolumns \\
			23.4 & \othercolumns \\
			33.1 & \othercolumns \\
			\hline
		\end{tabular}
	};

	\node (T3) at (\targetX, -\targetY * \OffsetY) {
		\begin{tabular}{|C{\TargetColWidth}|c|c|c|}
			\hline
			\cellcolor{focuscolor}\textbf{CEO} & \multicolumn{\otherColumnsNum}{>{\columncolor{graycol}}c|}{\otherColumnsHeader} \\
			\hline
			Satya Nadella & \othercolumns \\
			Ramon Laguarta & \othercolumns \\
			Koji Sato & \othercolumns \\
			\hline
		\end{tabular}
	};
	
	\draw[-latex, line width=1.5pt] (S1) -- (T1);
	\draw[-latex, line width=1.5pt] (S2) -- (T2);
	\draw[-latex, line width=1.5pt] (S3) -- (T3);

\setlength\extrarowheight{0pt}

\end{tikzpicture}
\vspace{-30pt}
		\caption{Example input tables with formatting mismatch}
		\label{fig:etables}
	\end{figure}

	The problem can be formulated as given a small set of matched rows between source and target tables, we want to identify mappings that transform source values into target values, enabling an equi-join between the two tables. 
	This paper explores two key research questions in this context: (1) Can higher performance be achieved in transforming tables that exhibit a wide range of syntactic and semantic patterns (as illustrated in Figure~\ref{fig:etables})? (2) Can the transformation process be made explainable?
	Aligned with prior work in the literature~\cite{dtt, gxjoin, autojoin, FlashFill}, this work primarily focuses on transformations applied to key or join columns. However, the proposed approach has the potential to be extended to include additional columns or more complex settings.

	\textbf{Challenges.} 
	There are a few critical challenges that must be addressed to tackle the problem.
	One major challenge is \textit{the vast multi-class search space} due to the diverse nature of mismatches, as illustrated in Figure~\ref{fig:etables}. 
	These mismatches can range from string transformations, such as converting full names to usernames, to numeric conversions, like pounds to kilograms. In more complex cases, transformations may even require external knowledge bases (for example, linking a company name to its CEO). An effective solution must not only handle all of these classes but also navigate the vast search space of potential transformations within each class to identify the most appropriate mappings.
	Another challenge arises from \textit{the limited coverage of individual transformations}. A single transformation may only be applicable to a subset of the rows, as demonstrated in the first table in Figure~\ref{fig:etables}, where entries containing middle names are transformed differently than those without. A comprehensive framework must identify multiple transformations and develop rules to apply the appropriate transformation based on observed data patterns.
	Additionally, \textit{interpretability} is a crucial requirement for 
	many application areas, particularly those involving sensitive or high-stakes data. Users need to understand and verify the logic behind each transformation to ensure it meets their requirements. Furthermore, the ability to adjust or refine these mappings fosters trust and transparency in data integration processes.

	\textbf{Limitations of existing approaches.} 
	Existing methods for addressing table transformations can be broadly categorized into (1) methods that learn a matching function for a given pair of values or records, and (2) methods that generate mapping rules between source and target tables~\cite{dtt, gxjoin, icde, autojoin, FlashFill}.
	The first group includes techniques focused on learning similarity functions~\cite{auto-fuzzy-join, MassJoin} or mapping functions~\cite{semajoin, wang2017synthesizing}, as well as approaches for matching entity records that refer to the same real-world objects but differ in format or representation~\cite{akbarian2022probing, Li2020:Deep, auto-em}. 
	These approaches typically rely on machine learning models or (semi-)manually defined similarity metrics to rank and select the best candidate for each corresponding value. While effective for addressing formatting mismatches in joining tabular data, they lack the ability to generate explicit mappings to transform the source formatting into the target. This limits their flexibility in modifying transformations, may lead to non-interpretable functions, and restricts their use primarily to fuzzy joins.
	The second group of approaches focuses on generating mapping rules.
	Many of these studies are limited to string-based transformations and rely on an exhaustive search of the parameter space to find the appropriate mapping. Some methods are constrained to using a single transformation function per table~\cite{autojoin,BlinkFill,FlashFill,FlashFill2}, while others depend entirely on retrieving functions from external sources~\cite{TDE,TDE2}. Also, some approaches employ pretrained language models, which return mappings that are not interpretable~\cite{dtt}.
	
	\textbf{Our approach.} 
	We introduce TabulaX, a novel data integration framework that leverages Large Language Models (LLMs) for multi-class column-level tabular transformations. Similar to existing approaches, TabulaX transforms the entity or join column in the source table to its corresponding representation in the target. However, the mapping functions generated by our framework cover a much broader range of transformations, including textual, numeric, algorithmic, and those requiring external knowledge.
	A key feature of TabulaX lies in the interpretability and usability of its transformations, which are generated as numeric functions or programming language code, unlike some approaches in the literature that rely on their own domain specific languages~\cite{BlinkFill,FlashFill,FlashFill2}.
	Our experimental evaluation reveals that TabulaX outperforms existing state-of-the-art approaches in terms of accuracy, supports a broader class of transformations, and delivers mappings that can be efficiently applied to large-scale tables and datasets. 
	
	\textbf{Contributions.} 
	Our contributions can be summarized as follows:
	(1) We propose TabulaX, a new framework for Multi-class Tabular Transformations, capable of handling a broad range of transformations---including textual, numeric, algorithmic, and those requiring external knowledge.
	(2) We introduce a classification mechanism that enables TabulaX to effectively apply tailored transformation strategies to diverse types of data mismatches.
	(3) We develop methods for generating human-interpretable transformation functions, such as numeric formulas and executable code, which improve transparency and allow users to understand, verify, and modify the mappings.
	(4) Through extensive experiments on real-world datasets from various domains, we demonstrate that TabulaX outperforms existing state-of-the-art approaches in terms of accuracy and supports a broader class of transformations.
	(5) We publicly release the TabulaX framework, source code, and datasets to support reproducability and foster further research\footnote{https://github.com/arashdn/TabulaX}.

	\section{Related Works}
	We organize related work into three main areas: (1) example-driven table transformations, (2) the use of large language models for tabular data, and (3) techniques for integrating tabular datasets.

	\subsection{Example-Driven Table Transformations} 
	This line of work, closely aligns with ours and has been an active area of research in recent years~\cite{dtt,icde,autojoin,BlinkFill,FlashFill,FlashFill2}. Early approaches such as FlashFill~\cite{FlashFill} and BlinkFill~\cite{BlinkFill} targeted spreadsheet data using substring-based transformations derived from user examples. 
	Auto-join~\cite{autojoin} improves robustness by employing predefined string transformation units and a recursive backtracking algorithm to partition and transform noisy input.  CST~\cite{icde} builds on this by extracting common text fragments  as transformation skeletons, applying row-level transformations, and ranking them by coverage. GXJoin~\cite{gxjoin} generalizes these transformation units, improving both coverage and noise tolerance.

	While CST and GXJoin improve efficiency and noise handling, they remain limited to a narrow set of predefined substring-based operations and require long textual overlaps.
	To address these constraints, DTT~\cite{dtt} introduces a fine-tuned language model for example-driven transformations. This approach eliminates the need for predefined functions or pruning-based searches but relies  on synthetic training data and produces transformations in a non-interpretable latent space. Our work addresses these problems by introducing a novel class-aware LLM-based framework capable of generating interpretable, human-readable transformation functions, bridging expressiveness, generalization, and transparency.
	
	Some structural transformation methods, such as Foofah~\cite{foofah} and Explain-Da-V~\cite{shraga2023explainableDAV}, focus on reshaping entire tables. While they target cell- and grid-level alignment for schema matching, our column-level mappings can complement these approaches by handling diverse value-level mismatches in integration tasks.
	
	\subsection{Large Language Models for Tabular Data}
	Early pretrained models such as T5~\cite{t5}, BERT~\cite{bert}, and GPT-3~\cite{gpt3} have inspired the development of tabular-specific LLMs 
	such as TaBERT~\cite{tabert}, TURL~\cite{turl}, and TABBIE~\cite{tabbie}. These encoder-based models are better suited for discriminative tasks like entity matching~\cite{DITTO,akbarian2022probing} and QA~\cite{chen2020open,tabert,tabbie}, while models like RPT~\cite{rpt} and ByT5~\cite{byt5} support generative tasks such as data to text~\cite{parikh-etal-2020-totto,kale-rastogi-2020-text,thorne2021natural}. Tabular LMs typically have less than 500M parameters and rely on fine-tuning for task adaptation. DTT~\cite{dtt}, for instance, is built on fine-tuned ByT5~\cite{byt5} for table joinability.
	
	Recent advances in LLMs such as ChatGPT\footnote{https://chatgpt.com}, Gemini\footnote{https://gemini.google.com}, and Llama 3~\cite{llama} have enabled prompt-based methods that require minimal or no fine-tuning. Techniques such as self-consistency~\cite{wang2022self} and chain-of-thought prompting~\cite{CoT} improve LLM reasoning and performance by guiding generation through structured prompts.
	Some studies apply LLMs to tabular tasks using  supervised fine-tuning~\cite{tablellama,tablegpt} and prompt-based approaches~\cite{sui2024table}, emphasizing the importance of table serialization~\cite{dtt,normtab,sui2024table}.
	These approaches are complementary to ours, improving how LLMs process and reason about tabular data.
	
	\subsection{Tabular Data Integration}
	Research focused on linking and integrating structured datasets is closely related to our goals. A comprehensive overview of foundational methods is provided by Doan et al.~\cite{doan2012principles}, while more recent developments, particularly in the context of data lakes and open data integration, are reviewed by Miller~et~al.~\cite{miller2018open} and Khatiwada~et~al.~\cite{khatiwada2022integrating}.
	Our work also connects with efforts to detect linkage points across datasets~\cite{hassanzadeh2013discovering}. Notably, when schema elements (e.g., \texttt{dbpedia:stockSymbol} and \texttt{dbpedia:stockTicker}) are interpreted as cell values, their alignment can be viewed as a form of cell-level transformation. This perspective supports the idea that schema-level and value-level integration can be unified under a single transformation framework, which our method aims to achieve through class-aware, interpretable mappings.

	\section{Multi-Class Transformation Discovery} 
	
	In this section, we provide a detailed formulation of the problem, a discussion of our assumptions and observations, and a classification of transformations based on their functions.

	\label{sec:problem_def}
	Our aim is to transform the values in a source table column into their corresponding values in a target column, guided by a small set of provided examples. We assume that both the source values to be transformed and user-provided examples are available, which helps narrow the scope to concentrate specifically on data transformation tasks~\cite{dtt,icde}. When such examples are not provided by users, methods such as unequal joining~\cite{auto-fuzzy-join,semajoin,wang2017synthesizing} or token-based example generation~\cite{icde,autojoin} can be employed to produce a set of examples, with the caveat that these automatically generated examples may include noise or invalid matches.

	Let $S=\{s_1, s_2, \ldots \}$ denote a set of values in the source table, and  $T=\{t_1, t_2, \ldots \}$ represent the corresponding values in the target, which may be partially or completely unavailable.  For a small subset $S^\prime \subset S$, let $E=\{(s_i,t_j) | s_i \in S^\prime\}$ denote a set of examples where target values are provided to guide the transformation process.
	We want to find a transformation function $f$ such that: 
	\begin{equation}
		\label{eq:function}
		f: S \rightarrow T | \forall s_j \in S^\prime ((s_j, f(s_j) \in E.
	\end{equation}
	Ideally, the function learned from $S^\prime$ should be applicable to any value in $S$, rather than being restricted only to $S^\prime$. 
	The function is expected to be human-interpretable, meaning that its output should be generated in a manner supported by transparent patterns. Additionally, it should be implementable using commonly utilized programming languages in organizational settings. 
	Moreover, given that many data lakes and tables sourced from third parties often lack schema information, we operate under the assumption that no metadata or schema is available for the framework to benefit. Accordingly, transformations will be extracted only using the values in the individual cells.
	
	As an example, consider the middle row in Figure~\ref{fig:etables}, where \verb|S = {2, 51.5, 73}| and \verb|T = {0.9, 23.4, 33.1}|. We provide the set of guiding examples as \verb|E = {(2, 0.9), (51.5, 23.4)}|. The aim is to find a transformation function that maps \verb|2| to \verb|0.9|, and \verb|51.5| to \verb|23.4|, thereby identifying a pattern. For instance, the function $f(x) = \text{round}(0.453 \times x)$ can transform any source value in $E$ into its corresponding target and is generalizable to all inputs in $S$.

	\begin{figure*}[tbp]
		\centering
		\includegraphics[width=1\linewidth]{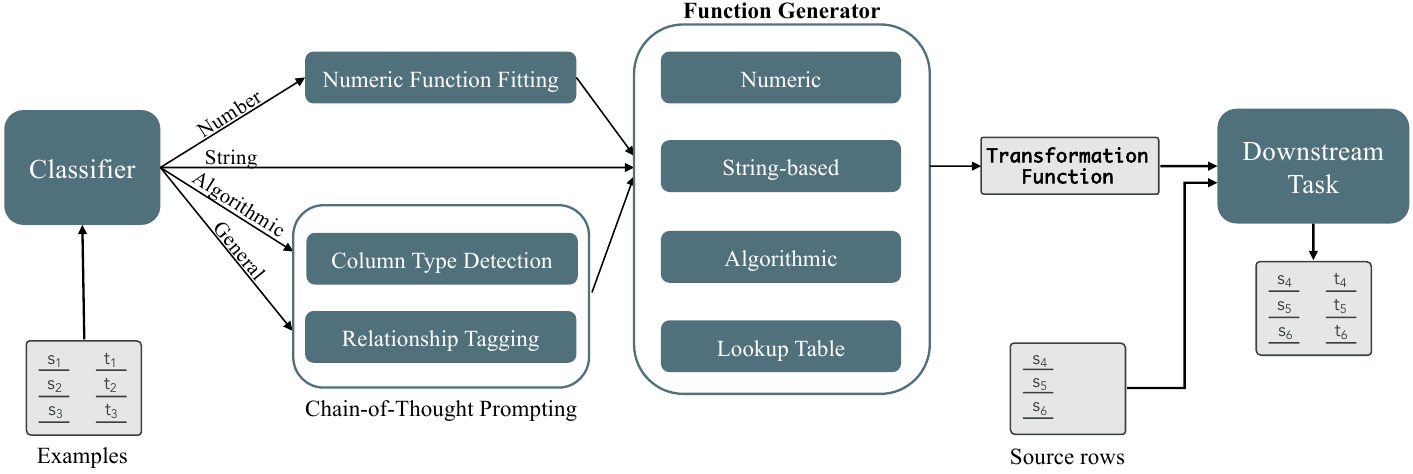}
		\caption{The architecture of our framework}
		\label{fig:framework}
	\end{figure*}

	As shown in Figure~\ref{fig:etables}, different input tables require distinct classes of transformations. For example, the top table can be transformed using a sequence of string manipulation functions, such as substring and split, while the middle table relies on a numerical transformation (specifically, a linear function). Transformation of the last table is not possible without external knowledge that is not present in either the source or target.
	
	Most existing approaches in the literature limit their scope when addressing this problem. Some methods focus only on string-based transformations~\cite{dtt,autojoin,FlashFill}, others only handle entity matching~\cite{auto-em,DITTO}, and some retrieve mapping functions from large code repositories~\cite{TDE,TDE2}. One may argue that recent commercially available LLMs, such as GPT and Gemini, could address these challenges. To evaluate the performance of LLMs in this context, we conducted an experiment using the GPT model\footnote{We used the \texttt{gpt-4o-2024-05-13} model from the OpenAI API for this experiment} on our benchmark consisting of diverse real-world tables. However, the out-of-the-box model struggled to recognize many textual patterns, and frequently produced hallucinated or irrelevant outputs in cases requiring external knowledge. Moreover, LLMs are not expected to handle numerical data well~\cite{chang2024survey}, and our experiments also confirmed that the model was unable to perform numerical transformations effectively.
	
	To address this issue, we propose a framework that first classifies the input mappings. Based on the identified class, appropriate actions and methods are then selected to facilitate the transformation of the tables. We define four distinct classes of input transformations:
	\begin{itemize}
		\item \textbf{String-based}: This class includes input columns that can be transformed using a series of string manipulation functions commonly available in programming frameworks, such as substring, concatenation, split, case conversion (upper/lowercase), etc. An example of this class is shown in the top row in Figure~\ref{fig:etables}.
		
		\item \textbf{Numerical}: This class encompasses input tables where both source and target values are rational numbers and exhibit an underlying mathematical relationship. An example is the conversion from pounds to kilograms illustrated in Figure~\ref{fig:etables}. It is important to note that nominal numbers, such as postal codes or ISBNs, do not fall in this category.
		
		\item \textbf{Algorithmic}: This class includes input tables where an algorithm exists to transform source values into target values, but the transformation goes beyond numerical functions or sequences of string operations. Examples include transformation such as Gregorian to Hijri dates, characters to ASCII codes, or binary to hexadecimal numbers.
		
		\item \textbf{General}: This class includes any table transformations that require external knowledge sources or do not fit into the previous three categories. Examples include mapping a company name to its CEO, converting area codes to countries, or translating English words to French.
	\end{itemize}
	
	In the following sections, we will discuss the classification process and the specific actions required for each class of input tables.

	\section{Approach}
	\label{sec:approach}
	Figure~\ref{fig:framework} depicts the architecture of TabulaX, our proposed framework for tabular data transformation and join. The first component is a classifier that assigns a transformation class from the aforementioned set of classes. Once classified, the tables are routed to the corresponding mapping modules, which generate the appropriate transformation functions. These transformations are then utilized by the downstream task component to produce end-to-end results. While this study focuses on unequal joins as the downstream task, the generated transformations can also support other applications, such as detecting anomalies or outliers by identifying deviations from the defined mappings, imputing missing values, etc.
	The remainder of this section details each component of the framework.
	
	\subsection{Input Data and Serialization}
	\label{sec:input_data}

	Multiple components in our approach interact with LLMs, and most LLMs accept input only as a sequence of tokens, hence a critical step is serializing the input tables into a linear sequence of text. This sequence is then tokenized and processed by the LLM. Most LLMs are trained on vast, diverse datasets from web pages and are familiar with common notations used to represent sets and pairs in mathematics and computer science. Leveraging this, we serialize the example pairs in a way that closely aligns with these conventions.
	
	Specifically, each example is serialized as \verb|("s" -> "t")|, where \verb|s| and \verb|t| represent the source and target values, respectively. To ensure generality and with the assumption that schema and metadata may not be available, all values are enclosed in quotation marks, regardless of their cell or column data types. We use the \verb|->| token to indicate the direction from source to target, instead of a comma, which is more commonly used for separating elements in pairs or tuples. This decision is supported by our observations that when a comma is used to separate tuple elements, LLMs, particularly GPT-4o, may fail to recognize the directionality from source to target, leading to erroneous predictions. Finally, the serialized examples are concatenated using commas to form the complete serialized list.
	
	For instance, consider the first two rows from the bottom row in Figure~\ref{fig:etables} as the set of provided examples. This set will be serialized as: \\ \texttt{("Microsoft" -> "Satya Nadella"),( "PepsiCo" -> "Ramon Laguarta")}.\\ This serialization will be utilized throughout the framework whenever an LLM needs to process the example set. Since many LLMs impose a limit on input length, if the size of the example set exceeds the model's context window, a sampling step is applied during serialization to reduce the input length.

	\subsection{Classifier}
	Our experiments with various LLMs indicate that even large, widely-used commercial models such as GPT-4o struggle to effectively transform numeric values, recognize all textual patterns, and generate lookup tables without prompts that are tailored for each class. To address this, the first component of our framework is a classifier that determines the subsequent components to which the input tables will be directed. To achieve this, we use a general-purpose LLM. LLMs, trained on vast and diverse datasets, can handle a wide range of tasks by utilizing one or a few training examples (i.e., shots)~\cite{gpt3,sahoo2024systematic} in their prompts, known as in-context learning. Inspired by this idea, we developed a prompt template for classifying input types. The structure of the prompt and the given examples (shots) may significantly affect the quality of the classification, which led us to try 3-5 prompts before selecting one.
	Our classifier's LLM prompt\footnote{The prompt template, along with all other prompting techniques and templates used in this study, are available in our publicly-available code repository.} starts by defining the task, followed by a list of possible classes. Each class is then explained in detail, and representative examples are provided at the end.

	Once the input class is identified, we propose a tailored method that best suits the transformation of values in that class. 
	Recognizing that LLMs face challenges with numerical mappings, our framework avoids relying on LLMs for this class. 
	It also uses Chain-of-Thought prompting~\cite{CoT} for advanced algorithmic and general classes to enhance the model's reasoning capabilities. In what follows, we will describe how transformations are generated for each class in more detail.

	\subsection{Numerical Transformations}
	When the input table is labeled as numerical, it is passed to the Numerical Function Fitting component, which is responsible for fitting a numerical function $f_n$ (i.e., curve fitting) based on the input examples. The framework attempts to fit curves using the following types of functions:
	\begin{itemize}
		\item Linear: $f(x) = ax + b$,
		\item Polynomial: $f(x) = ax^2 + bx + c$,
		\item Exponential: $f(x) = a \text{\space} exp(bx)$,
		\item Rational: $f(x) = \frac{ax + b}{x + c}$.
	\end{itemize}
	The parameters $a$, $b$, and $c$ are estimated during the curve fitting process using the Levenberg–Marquardt algorithm~\cite{ranganathan2004levenberg}. The framework then calculates the Mean Square Error (MSE) between the predicted and actual target values in the provided examples, and the function with the lowest MSE is selected as the mapping function, denoted as $f_n(x)$.
	The list of functions is not limited to those mentioned above; any function supported by the curve fitting algorithm can be used.
	
	Once the best-fitting function $f_n(x)$ is selected, it is passed to the Numerical Function Generator component, which generates the corresponding transformation as a function in a programming language---specifically Python, in our setup.
	
	\subsection{String-Based Transformations}
	Table transformations using string primitives and regular expressions have been widely studied in the literature~\cite{gxjoin,icde,autojoin,BlinkFill,FlashFill}, often using a Programming-By-Example (PBE) approach. While any of these methods could be incorporated as our String-Based Function Generator, we propose a novel approach by leveraging LLMs, which outperform state-of-the-art techniques. Recent studies suggest that LLMs perform well in code generation~\cite{wang2023review,liu2024empirical,nejjar2023llms}, and our approach also utilizes LLMs to generate transformation functions for string-based inputs.

	Specifically, the model is prompted to create a Python function that takes a string as input and reformats it to produce the expected output. In this case, the input examples are not presented using the set-like notation from Section~\ref{sec:input_data}. Instead, we format them as test cases: \texttt{Input: $s_i$, Expected output: $t_i$}, where $s_i$ and $t_i$ represent arbitrary source and target samples, respectively. This format is common on websites and forums, which are primary sources for common LLMs training data. As a result, LLMs better understand the task using this format, as demonstrated in our experiments, leading to improved code generation compared to the set-like notation.

	Additionally, the model is instructed not to use any external library calls but is free to utilize any functions available in the standard Python library. This approach covers a broader range of string-based transformations compared to methods that rely on a limited set of predefined functions~\cite{gxjoin,icde,autojoin} or those using substring extraction via regular expressions~\cite{BlinkFill,FlashFill,FlashFill2}. Finally, the generated function is verified by a syntax checker before being used as the transformation function.

	\subsection{Algorithmic Transformations}
	Algorithmic transformations are more complex compared to numerical and string-based transformations. Our experiments indicate that applying the same approach used for string-based transformations is ineffective here, mainly due to the limitations of LLMs in handling complex reasoning. In many cases, the model fails to identify the correct transformation pattern and may hallucinate and generate code that does not align with the given examples.

	Studies suggests that complex reasoning in LLMs can be improved through intermediate reasoning steps~\cite{CoT,ToT}, known as Chain-of-Thought (CoT) reasoning. Leveraging this approach, we break down the task of generating algorithmic transformations into two simpler steps: (1)  \textit{Relationship Tagging}, which identifies the conversion type or the relationship between the source and target, and (2) \textit{Algorithmic Function Generator}, which produces a Python function to execute the transformation based on the identified relationship.

	For tables requiring algorithmic transformations, the input is first processed by the relationship extractor, where the LLM is prompted to identify the relationship between the source and target columns. The relationship is formatted as \texttt{[type of source column] to [type of target column]}. For example, given the input \texttt{("2024/09/05" -> "1403/06/16"), ("1886/06/27" -> "1265/04/06")}, the model is expected to return \texttt{Gregorian date to Jalali (Solar Hijri) date}. This transformation is included in the training examples provided for in-context learning, alongside others, such as email to domain transformation.
	The detected relationship and input tables are then passed to the algorithmic function generator component, which is structured similarly to the string-based function generator.
	In this step, the model generates a Python function based on the identified relationship, using the same format and verification as the string-based component.

	\subsection{General Transformations}
	General transformations are the most complex task, as they often require external knowledge not present in the provided examples. We adopt a similar approach to the one used for algorithmic transformations, supported by Chain-of-Thought prompting, and break the problem into two subtasks. The first step is similar to the relationship tagging component in the algorithmic transformer, where the types of source and target columns are detected. 
	The second step generates a transformation by leveraging a lookup table (i.e., a structured mapping between source and target values) as a bridge. There are multiple ways to obtain this table. One approach is to retrieve it from external repositories (e.g., structured databases, web tables, or open knowledge graphs), which requires extensive indexing and specialized retrieval algorithms~\cite{TDE,TDE2,semajoin}. In this work, however, we leverage LLMs to dynamically construct the lookup table, allowing for greater flexibility and adaptability.

	When input tables are labeled for general transformation, they are passed to the \textit{Column Type Detection} component, where the LLM is prompted to identify the type of the source and target columns, formatted as \texttt{[type of source column] to [type of target column]}. While the task shares some similarities with algorithmic transformer, the two training examples used in this context are selected to better match the general transformation scenario. For instance, given \texttt{Data: ("AGC" -> "United States"), ("YYZ" -> "Canada"),} the expected relationship is \texttt{Airport Code to Country} and for \texttt{Data: ("gallon" -> "liter"), ("inch" -> "centimeter"),} the relationship will be \texttt{Imperial Units to Metric Unit}.

	As demonstrated in these examples, developing a deterministic algorithm for such cases is infeasible, as the transformations rely on external knowledge. Consequently, we leverage the LLM's learned knowledge to generate a lookup function that directly prompts the model to predict the target value for a given source value, based on the extracted column types and input examples.

	\begin{figure*}[tbp]
		\centering
		\includegraphics[width=1\linewidth]{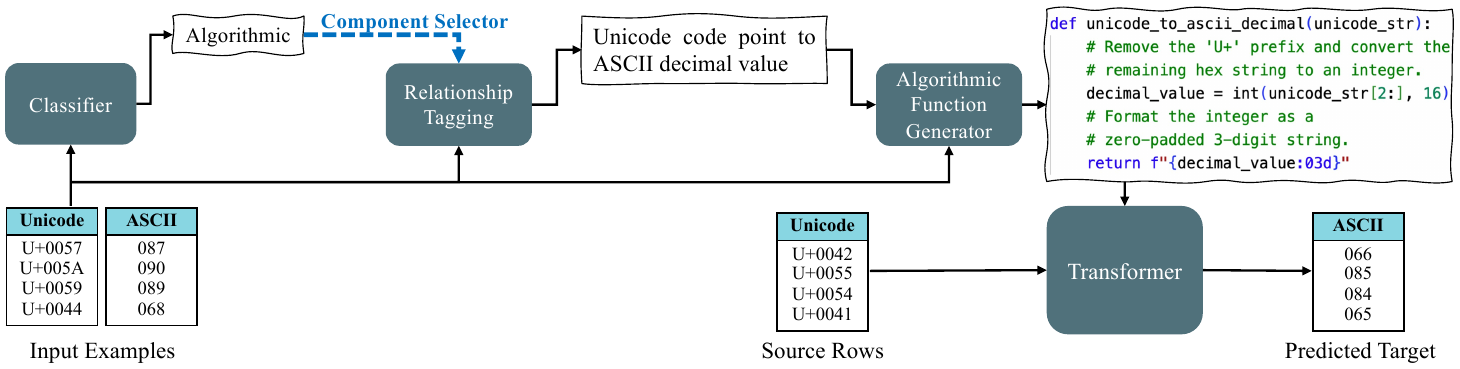}
		\caption{The steps of processing an arbitrary input table}
		\label{fig:tr_example}
	\end{figure*}
	
	\par\noindent
	\textbf{Limitations of general transformations.} 
	While our task breakdown and other steps significantly improve the model's performance compared to state-of-the-art baselines, there are certain limitations. First, the external knowledge used by the model is inherently constrained by its training data. Despite the extensive training of modern LLMs on large datasets from web pages and public knowledge bases, they may still lack domain-specific knowledge, particularly in private sectors where similar data is not publicly available. Additionally, the transformation function in this category primarily relies on LLM predictions, which are not human-interpretable. Unlike other transformation types where the mapping function is generated based on a single prompt using a small set of example rows---independent of the source column size---general transformations may require an LLM call for each row, particularly if no batching strategy is applied. While this may raise scalability concerns, it is worth noting that these LLM calls are independent and can be executed in parallel to deal with performance bottlenecks.

	Moreover, LLMs are susceptible to hallucinations, and unlike the numeric, string, or algorithmic transformers, where the output is an explicit algorithm that can be automatically or manually verified, hallucinations in this case may occur and remain undetected. In sensitive environments, one solution is to implement guardrails that verify the model's output using domain knowledge or compare it against available target values. This component could also be replaced with a retrieval-based approach to access relevant information from public knowledge bases or private data sources. For example, the system could be integrated with a table retrieval method that retrieves a bridge or cross table between the source and target~\cite{DataXFormer,semajoin,wang2017synthesizing}, or by employing entity matching techniques~\cite{auto-em,akbarian2022probing,Li2020:Deep}. These areas remain open for future works.

	\subsection{Example Walkthrough: Generating a Transformation}
	Figure~\ref{fig:tr_example} illustrates an end-to-end example of the framework in action. In this scenario, the source values represent Unicode code points of certain characters, while the target values correspond to their ASCII decimal representations. Four examples are provided to guide the transformation generation process. The examples are first passed to the classifier, which identifies the input type as an algorithmic transformation. Based on this classification, the framework invokes the algorithmic transformation process, beginning with the relationship tagging component. The model extracts the relationship as "Unicode code point to ASCII decimal value," which is then passed, along with the input examples, to the function generator component. The output of this step is a transformation expressed as human-interpretable Python code. This code is subsequently executed on the source rows to generate the predicted target values.
	
	\subsection{Integration with Downstream Tasks}
	\label{sec:downstream_tasks}
	The transformation functions generated by TabulaX can be directly applied to data analysis or used to support downstream tasks. Tabular transformations have been used for a variety of tasks, including auto-completion and auto-filling of spreadsheets~\cite{FlashFill,BlinkFill}, predicting missing values, error correction~\cite{he2016interactive}, and joining tables~\cite{gxjoin,autojoin}. In this section, we focus specifically on the task of joining heterogeneous tables, while the generalization of this framework for other downstream applications are left for future works.

	Consider a scenario where a source table $S$ and a target table $T$ must be joined on semantically equivalent columns, guided by a set of examples mappings $E$. While these columns may differ in format, they represent the same real-world entities or concepts. TabulaX leverages \( E \) to learn a transformation function, which is applied to each entry \( s_i \in S \) to produce a transformed value \( \hat{t}_i \). This results in a candidate set \( \hat{T} \), aligned with the format of \( T \). Treating \( \hat{T} \) as an intermediate column, an equality join with \( T \) can be used to complete the operation.

	However, strict equality joins are sensitive to minor discrepancies between predicted and actual values. Unlike tasks such as auto-completion, where exact matches are essential, join operations can often tolerate minor mismatches. When a one-to-one relationship is expected (e.g., in primary-foreign key joins), the join can instead be performed by selecting the closest match in \( T \) for each \( \hat{t}_i \), using string similarity metrics such as edit distance. Formally, the best match \( o_i \) for a source value \( s_i \) is given by:
	\begin{equation}
		o_i = \underset{t_j \in T}{\mathrm{argmin }} \text{\space} edit\_dist(\hat{t}_i, t_j). 
	\end{equation}
	This method can be generalized to handle one-to-many or many-to-many joins by defining upper and lower bounds on allowable edit distances. Edit-distance-based matching offers a transparent and targeted solution for handling small inconsistencies, though it is not intended as a full fuzzy matching algorithm.
	
	For numeric transformations, edit distance is not appropriate. Instead, we use absolute numerical differences to measure proximity between predicted and actual values. This maintains consistency with the goal of interpretable and context-sensitive alignment while accounting for datatype-specific variation.
	Edit-distance-based matching is a simple yet effective technique, specifically used to address minor discrepancies between the predicted and target values, while maintaining transparency. It is not intended to perform a comprehensive fuzzy join but rather serves as a targeted approach to enhance matching accuracy in the presence of slight inconsistencies.
	Unlike the other classes, however, edit-distance-based matching is not applicable on numeric transformations; instead, we employ absolute numerical distance to assess the closeness of the transformed values to their targets.
	
	\section{Experiments and Analysis}
	This section presents a comprehensive evaluation of our framework, examining its performance across different configurations and benchmarking it against established baselines.

	\subsection{Benchmark Datasets}
	\label{sec:datasets}
	To evaluate the performance of our proposed method and compare it with current state-of-the-art baselines, we employ four real-world datasets encompassing various input classes. Detailed descriptions of each dataset are provided below.
	
	\noindent 
	\textbf{Web Tables Dataset (WT):} Initially introduced by Zhu~et~al.~\cite{autojoin} and subsequently utilized as a benchmark by Nobari~et~al.~\cite{dtt,icde}, this dataset consists of 31 pairs of tables spanning 17 distinct topics. Each table averages 92.13 rows with an input length of approximately 31 characters. The tables were extracted from Google Fusion Tables by selecting those that appeared in the same query results but had different formats. This benchmark involves string-based transformations and includes natural noise and inconsistencies. Not all entities can be transformed using traditional primitive string-based transformations, making this dataset relatively challenging string-based input~\cite{icde,dtt}.
	
	\noindent
	\textbf{Spreadsheet Dataset (SS):} This dataset comprises 108 pairs of tables sourced from the Microsoft Excel product team and user help forums, focusing on users' data cleaning issues. The tables represent spreadsheet pages that convey the same information in different formats. It includes public benchmarks from FlashFill~\cite{FlashFill} and BlinkFill~\cite{BlinkFill}, and was featured in the 2016 Syntax-Guided Synthesis Competition (SyGuS-Comp)~\cite{sygus2016}. On average, each table contains 34.43 rows and has an input source length of 19 characters. Compared to the WT dataset, while SS dataset is also string-based, it features considerably less noise and inconsistency.
	
	\noindent
	\textbf{Table Transformation (TT):} 
	Introduced by He~et~al~\cite{TDE}, this dataset was developed to support the task of retrieving data transformations from public code bases. It contains 230 pairs of tables, each averaging 8 rows, and covers a wide range of transformation types, including 102 algorithmic, 57 numeric, 68 string-based, and 3 general transformations. Compared to the WT and SS datasets, TT features more complex transformation logic, often involving numerical computations and algorithmic operations in addition to advanced string manipulations.

	\noindent
	\textbf{Knowledge Base Web Tables (KBWT):} Introduced by Abedjan~et~al.~\cite{DataXFormer}, this dataset primarily consists of tables extracted from a Knowledge Base (KB), requiring semantic transformations and additional KB information. For our evaluation, we selected single-column tasks from this dataset, totaling 81 pairs of tables. Each table averages 113 rows with an input source length of 13 characters. This dataset differs notably from the WT and SS benchmarks, which focus mainly on textual transformations. Among the tables in the KBWT benchmark, three are numeric, four are algorithmic, and the rest are general transformations.

	\subsection{Experimental Setup and Evaluation Metrics}
	Our framework, TabulaX, follows an example-driven approach, taking $n$ input examples to guide the transformation process. TabulaX is designed to allow any general-purpose LLM to serve as the inference model within the framework. We experiment with various numbers of input examples and multiple LLMs to provide a detailed analysis of our framework's performance. Where the value of $n$ is not explicitly mentioned, it is set to $5$ for the WT, SS, and KBWT datasets. For the TT dataset, $n$ is set to $3$ due to the relatively small number of rows in most of its tables. The default LLM used in the framework is OpenAI GPT-4o model, specifically, \texttt{gpt-4o-2024-05-13} version.
	
	For each transformation class, we tested 3–5 prompt variants and selected the best-performing one for our experiments. In the string-based and algorithmic transformations, where the output domain is relatively constrained and the final task is code generation, we observed that changes in prompt had only a subtle effect on performance. In contrast, general transformations involve more diverse domains, and prompt variations had a somewhat greater impact. For this class, we selected the most general prompt that yielded the best overall results. Overall, performance remained stable across prompts that followed the guidelines described in Section~\ref{sec:approach}, especially for larger models like GPT-4o.

	The focus of this study is the heterogeneous join as the downstream task, and as described in Section~\ref{sec:downstream_tasks}, the prediction is obtained by calculating the edit distance between the predicted and target values. While this is the default approach in our framework, we also conduct some experiments using exact matching to showcase our model's generalizability to other downstream tasks where target values may be unavailable or edit distance may not be preferred.
	
	Join performance is evaluated based on precision, recall, and F1-score, where precision measures the fraction of correct predictions that join with the target, recall represents the fraction of source rows that are correctly mapped, and F1-score is calculated by taking the harmonic mean of precision and recall. It is important to recognize that not all source rows may be mapped to a target value for reasons such as generating an invalid function, returning empty output, or runtime exception.
	In addition to those metrics, we also report the Average Edit Distance (AED) and Average Normalized Edit Distance (ANED), which indicates how much a prediction deviates from its target. ANED is normalized by the target length, facilitating comparisons between datasets and tables with different lengths.
	Each dataset's reported metrics are averaged across all tables within that dataset.

	\begin{table*}[htbp]
		\centering
		\setlength\extrarowheight{1pt}
		\setlength{\tabcolsep}{7pt}
		\caption{Classification performance of various LLMs across transformation classes}
		\begin{tabular}{|l|c||c|c|c||c|c|c||c|c|c|}
			\toprule
			\multicolumn{2}{|c||}{} & \multicolumn{3}{c||}{GPT-4o} & \multicolumn{3}{c||}{GPT-4o-mini} & \multicolumn{3}{c|}{LLaMA 3.1} \\
			\midrule
			Class & Support & P     & R     & F     & P     & R     & F     & P     & R     & F \\
			\midrule
			String & 207   & 0.92  & 0.85  & \textbf{0.88} & 0.89  & 0.74  & 0.81  & 0.72  & 0.91  & 0.80 \\
			Numbers & 60    & 0.78  & 1.00  & \textbf{0.88} & 0.67  & 1.00  & 0.81  & 0.65  & 1.00  & 0.79 \\
			Algorithmic & 105   & 0.84  & 0.70  & \textbf{0.76} & 0.88  & 0.56  & 0.69  & 0.88  & 0.20  & 0.33 \\
			General & 76    & 0.78  & 0.96  & \textbf{0.86} & 0.59  & 0.92  & 0.72  & 0.53  & 0.49  & 0.51 \\
			\midrule
			Macro avg & 448   & 0.83  & 0.88  & \textbf{0.84} & 0.76  & 0.81  & 0.75  & 0.69  & 0.65  & 0.61 \\
			Micro avg & 448   & 0.86  & 0.85  & \textbf{0.85} & 0.81  & 0.77  & 0.76  & 0.71  & 0.68  & 0.64 \\
			\bottomrule
		\end{tabular}%
		\label{tab:classifier}%
	\end{table*}%
	
	\subsection{Performance of classification}

	Table~\ref{tab:classifier} presents the performance of our classification module using three different LLMs: GPT-4o, GPT-4o-mini, and LLaMA 3.1 8B. In this table, support refers to the number of tables in each transformation class. To measure performance, we use common classification metrics: precision (P), recall (R), and F1-score (F). 
	These metrics are calculated across four transformation classes: String, Numbers, Algorithmic, and General. The macro average is calculated as the simple average of the metrics across all classes, regardless of their support, while the micro average takes into account the weight of each class by considering the total number of instances. The goal is to assess how effectively each model can categorize input tables into the correct transformation class, which is crucial for directing them to the appropriate mapping components in our framework.

	While the benchmark datasets provide source and target table pairs for evaluation, they do not include transformation type labels required for the classification task. To support this component, the transformation classes for all benchmark instances were manually annotated by the authors. During the annotation process, a conflict rate of approximately 8\% was observed, primarily in distinguishing between algorithmic and string-based transformations. These disagreements were resolved through discussion and consensus to ensure consistency and accuracy in the labels.

	As expected, GPT-4o outperforms the other models across all classes, achieving the highest macro and micro average scores. This superior performance can be attributed to its larger size and its training focused more on understanding complex tasks. All models achieve perfect recall for the numerical class, indicating that they reliably detect numerical transformations. However, GPT-4o still outperforms the others in precision, suggesting fewer misclassifications, such as incorrectly labeling nominal numbers (like zip codes) as numerical transformations or failing to recognize when only one side of the transformation involves numerical data. As a result, they achieve high recall, while their precision is not perfect.

	Another important observation is the tendency of the models to misclassify certain String transformations as General or Algorithmic. This misclassification often occurs in cases requiring extensive edit operations or the addition of literal text. For example, when transforming usernames into email addresses by appending a domain name, these models sometimes label the task as General instead of String. They assume that adding a domain requires external knowledge rather than recognizing it as a fixed literal addition. This suggests that LLMS, especially the smaller ones, may struggle with more complex textual transformations that involve fixed patterns or extensive edits.

	Overall, GPT-4o demonstrates sufficient capability for our classification tasks, emphasizing the importance of choosing a model complex enough to prevent misclassifications that could impact the subsequent mapping components in our framework. In certain domains, the transformation classes may already be established or can be determined by domain experts, eliminating the need for this classification component altogether.

	\subsection{Performance Compared to baseline models}
	
	In this section, we evaluate the performance of our framework, TabulaX, on the end-to-end task of heterogeneous or unequal table joins. This task simulates a common scenario where source and target columns reside in different tables that need to be joined, but the values in these columns are formatted differently.

	\begin{table*}[htbp]
		\centering
		\setlength\extrarowheight{1.8pt}
		\caption{Performance compared to the baselines}
		\begin{tabular}{|l||c|c|c||c|c|c||c|c|c||c|c|c||>{}c|>{}c|>{}c|}
			\toprule
			& \multicolumn{3}{c||}{TabulaX} & \multicolumn{3}{c||}{DTT} & \multicolumn{3}{c||}{GXJoin} & \multicolumn{3}{c||}{AFJ} & \multicolumn{3}{c|}{Explain-Da-V} \\
			\midrule
			Dataset & P     & R     & F     & P     & R     & F     & P     & R     & F     & P     & R     & F     & P     & R     & F \\
			\midrule
			WT    & 0.985 & 0.980 & \textbf{0.983} & 0.951 & 0.950 & 0.950 & 0.953 & 0.754 & 0.777 & 0.935 & 0.672 & 0.708 & 0.161 & 0.161 & 0.161 \\
			SS    & 0.955 & 0.949 & \textbf{0.952} & 0.954 & 0.952 & \textbf{0.952} & 0.997 & 0.787 & 0.810 & 0.943 & 0.662 & 0.691 & 0.163 & 0.163 & 0.163 \\
			TT    & 0.919 & 0.918 & \textbf{0.918} & 0.644 & 0.643 & 0.643 & 0.983 & 0.202 & 0.223 & 0.591 & 0.223 & 0.251 & 0.219 & 0.219 & 0.219 \\
			KBWT  & 0.722 & 0.532 & \textbf{0.567} & 0.276 & 0.248 & 0.254 & 0.962 & 0.081 & 0.083 & 0.749 & 0.067 & 0.093 & 0.038 & 0.038 & 0.038 \\
			\bottomrule
		\end{tabular}%
		\label{tab:baselines}%
	\end{table*}%
	
	To benchmark our approach, we compare TabulaX against four state-of-the-art baselines: Deep Tabular Transformer (DTT)~\cite{dtt}, GXJoin~\cite{gxjoin}, Auto-FuzzyJoin (AFJ)~\cite{auto-fuzzy-join}, and Explain-Da-V~\cite{shraga2023explainableDAV}.
	DTT leverages language models to generate outputs directly from provided examples, focusing on string transformations. GXJoin extends the CST method~\cite{icde} and exhaustively searches for general textual string-based transformations to facilitate table joinability. AFJ employs similarity functions to identify the most probable rows to join without generating explicit transformation functions. Explain-Da-V performs example-driven structural transformations by exploring a graph of intermediate table states with an A*-based heuristic search and type-aware operators, aiming to reshape source tables so they align with the target structure.
	
	Table~\ref{tab:baselines} summarizes the performance of our framework, TabulaX, and the baselines. As shown, TabulaX consistently achieves higher or comparable F1-scores across all datasets compared to these baselines. On datasets dominated by string-based transformations, such as WT and SS, the performance gap between TabulaX and DTT is relatively small, with DTT being the best-performing baseline in our setup. Specifically, TabulaX achieves an F1-score of 0.983 on the WT dataset, only marginally higher than DTT's 0.950. On the SS dataset, both methods perform almost identically, with F1-scores of 0.952. This similarity is expected since both models handle string transformations effectively, though TabulaX offers the added advantage of generating interpretable transformation functions. In contrast, DTT directly outputs target values without transparency, which can be a limitation in sensitive environments and scenarios requiring interpretability.
	The performance gap between TabulaX and the baselines increases on datasets involving more complex transformations. On the TT dataset, which includes a mix of string-based, numerical, and algorithmic transformations, TabulaX achieves an F1-score of 0.918, compared to 0.643 for DTT, 0.223 for GXJoin, and 0.219 for Explain-Da-V. The TDE approach~\cite{TDE} was proposed in the same work that introduced the TT dataset and reports a coverage of 72\% on it, whereas our method achieves a recall of 92\% on this benchmark. 
	On the KBWT dataset, which contains a higher proportion of general transformations requiring external knowledge, TabulaX reaches an F1-score of 0.567, whereas DTT scores only 0.254. The gap highlights TabulaX's versatility and its ability to generalize across diverse data types, a feature not shared by the string-based baselines. Additionally, TabulaX's interpretable transformations are valuable in understanding and verifying the mappings, unlike DTT's black-box approach.
	
	GXJoin, performs an exhaustive search in the string-based transformation space to generate interpretable transformations and is the best-performing state-of-the-art approach generating an interpretable mapping in our experiments. This large search space allows GXJoin to achieve high precision on all datasets, as it applies transformations without relying on edit distance-based matching. However, the recall is noticeably lower, indicating that it fails to transform a significant portion of the data. In contrast, TabulaX maintains a high recall by utilizing a broader set of transformation units and providing a clear mapping between input rows and transformations. Similar to other baselines, GXJoin is only limited to string-based transformations and is significantly outperformed by TabulaX and DTT on TT and KBWT datasets.
	Even without employing edit distance matching, as indicated in Table~\ref{tab:model_performance}, TabulaX outperforms GXJoin due to its ability to handle a wider variety of transformations and its use of LLMs to interpret and apply complex patterns. GXJoin's exhaustive search methodology, while thorough, can lead to scalability issues. 
	Moreover, while TabulaX provides a clear mapping between input rows and the applied transformations, GXJoin's exhaustive search does not offer such mapping, and it may apply multiple transformations without a straightforward way to trace back the exact steps for each row.
	
	In terms of runtime, for a table with $n$ rows, TabulaX requires only a single LLM call to generate transformations for string-based tables, two LLM calls (one for relationship tagging and another for transformation generation) for algorithmic tables, and $n+1$ LLM calls for general tables when lookups rely on LLM calls. Costs can be reduced if bridge tables are accessible from alternative sources. In contrast, DTT, while effective primarily for string-based tables, requires $k \times n$ LLM calls, where $k$ represents the number of self-consistency trials (set to 5 in the experiments). Both TabulaX (in the worst case) and DTT have a linear growth in the LLM calls with respect to input size, and their LLM calls can be executed in parallel if resources allow. On the other hand, GXJoin has quadratic growth in runtime due to its exhaustive search methodology. This approach also has more constraints on parallel execution, making GXJoin less scalable compared to TabulaX for larger datasets.

	AFJ relies on similarity functions to identify matching rows and does not generate explicit transformation functions, which limits its applicability in scenarios where understanding the transformation process is essential. Explain-Da-V, while capable of producing interpretable transformations, is primarily optimized for structural table reshaping. Its support for value-level transformations is limited, which hinders its performance in our setup. As a result, both methods underperform compared to TabulaX across all datasets, with particularly poor results on more complex benchmarks such as TT and KBWT. 
	
	Overall, our results demonstrate that TabulaX not only outperforms baselines on complex datasets but also provides interpretable transformation functions, making it a superior choice for heterogeneous joins across a wide range of tables.

	\begin{table*}[htbp]
		\setlength\extrarowheight{1.5pt}
		\centering
		\caption{The framework performance under various classification and matching methods }
		\begin{tabular}{|c|c||c|c|c|c|c||c|c|c|c|c|}
			\toprule
			\multirow{2}[4]{*}{\textbf{Classification}} & \multirow{2}[4]{*}{\textbf{Dataset}} & \multicolumn{5}{c||}{\textbf{Edit Distance Matching}} & \multicolumn{5}{c|}{\textbf{Exact Matching}} \\
			\cmidrule{3-12}          &       & \textbf{P} & \textbf{R} & \textbf{F1} & \textbf{AED} & \textbf{ANED} & \textbf{P} & \textbf{R} & \textbf{F1} & \textbf{AED} & \textbf{ANED} \\
			\midrule
			\midrule
			\multirow{4}[2]{*}{\begin{sideways}\textbf{LLM}\end{sideways}} & Web Tables (WT) & 0.985 & 0.980 & \textbf{0.983} & 1.501 & 0.068 & 1.000 & 0.729 & \textbf{0.816} & 5.497 & 0.269 \\
			& SpreadSheet (SS) & 0.955 & 0.949 & \textbf{0.952} & 1.944 & 0.094 & 1.000 & 0.818 & \textbf{0.828} & 4.879 & 0.180 \\
			& Table Transformation (TT) & 0.904 & 0.895 & 0.898 & 2.678 & 0.219 & 0.966 & 0.736 & \textbf{0.756} & 4.252 & 0.340 \\
			& KBWT  & 0.722 & 0.532 & 0.567 & 7.203 & 0.398 & 0.975 & 0.435 & 0.512 & 7.139 & 0.387 \\
			\midrule
			\midrule
			\multirow{4}[2]{*}{\begin{sideways}\textbf{Golden}\end{sideways}} & Web Tables (WT) & 0.975 & 0.971 & 0.973 & 2.223 & 0.094 & 1.000 & 0.690 & 0.770 & 6.362 & 0.308 \\
			& SpreadSheet (SS) & 0.953 & 0.949 & 0.951 & 2.132 & 0.104 & 1.000 & 0.798 & 0.809 & 5.270 & 0.202 \\
			& Table Transformation (TT) & 0.919 & 0.918 & \textbf{0.918} & 2.466 & 0.219 & 0.992 & 0.733 & \textbf{0.756} & 3.915 & 0.344 \\
			& KBWT  & 0.722 & 0.545 & \textbf{0.580} & 7.139 & 0.387 & 0.975 & 0.445 & \textbf{0.523} & 7.674 & 0.516 \\
			\bottomrule
		\end{tabular}%
		\label{tab:model_performance}
	\end{table*}

	\subsection{Impact of Matching Strategies}
	
	Table~\ref{tab:model_performance} presents a detailed comparison of the framework's performance under two different matching scenarios: edit-distance-based matching (left panel) and exact matching (right panel). The former matching strategy is  more suited for applications where a set of target values is available, such as in unequal table joins, where small variations between the generated and expected values can be tolerated. The latter, exact matching, is more appropriate for tasks like missing data imputation, where precise outputs are required.
	The top panel of the table reflects results when using the LLM-based classification model (GPT-4o in this set of experiments), while the bottom panel provides performance metrics when using the ground truth classification, referred to as the ``Golden'' classifier. This comparison allows us to assess the impact of the classification on the overall performance of the framework.

	An interesting observation is that the model performance on datasets consisting of string-based transformations (Web Tables and Spreadsheet) is slightly better when the classification is done via the LLM-based model compared to the golden classifier. For instance, on the Web Tables dataset using edit-distance matching, the F1-score with LLM-based classification is 0.983, marginally higher than the 0.973 achieved with the golden classifier. 
	This counterintuitive performance can be attributed to the LLM-based classifier occasionally mislabeling some difficult string-based transformations as general transformations. In these challenging cases, treating them as general transformations allows the model to generate the target values directly by leveraging the LLM's generative capabilities, without being constrained to produce a specific transformation function covering all input rows. While this approach can lead to better target predictions and higher recall, it sacrifices interpretability since the general transformations are not explicitly defined. In contrast, the golden classifier strictly labels these cases as string-based transformations, requiring the model to generate explicit transformation functions, which may be more difficult for complex patterns.

	Moreover, when exact matching is employed (right panel of the table), the model's precision increases to nearly perfect levels across all datasets and classification methods. This increase is expected, as outputs are only accepted if they exactly match the expected targets, significantly decreasing the possibility of false matches. However, this strict criterion leads to a reduction in recall, as the model misses rows where there are even minor discrepancies between the generated and expected values, such as differences in capitalization, punctuation, or minor formatting variations. In contrast, using edit-distance-based matching (left panel) allows minor differences between the predicted and target values, thereby considerably increasing recall and overall F1-score across all datasets. 
	The decrease in precision with edit-distance matching is minimal compared to the considerable increase in recall, resulting in a substantial boost in F1-score, which is a consistent trend across all datasets. Consequently, edit-distance matching emerges as a better choice when target values are available, as it enhances the model's overall performance by providing a better trade-off between precision and recall.
	
	This performance increase is more noticeable on string-based datasets, where minor inconsistencies are common. Specifically, on the Web Tables dataset under LLM-based classification, the F1-score improves from 0.816 with exact matching to 0.983 with edit-distance matching, representing a substantial increase of approximately 20.5\%. This demonstrates that allowing for slight variations in matching can significantly enhance the model's ability to correctly map source to target values in real-world, noisy datasets with inherent inconsistencies.
	On the other hand, on the TT and KBWT datasets, which include more algorithmic and general transformations and may rely on external knowledge, the improvement in F1-score when using edit-distance matching is less pronounced. For example, on the KBWT dataset, the F1-score under LLM-based classification improves from 0.512 with exact matching to 0.567 with edit-distance matching, an increase of approximately 10.7\%. This indicates that for general transformations, minor discrepancies are less probable to occur when the model is not constrained to a string-based transformation to cover all rows.

	\begin{table*}[htbp]
		\centering
		\caption{Performance Metrics per Transformation Class}
		\setlength\extrarowheight{1.5pt}
		\begin{tabular}{|l|c||c|c|c|c|c||c|c|c|c|c|}
			\toprule
			\multicolumn{1}{|c|}{\multirow{2}[3]{*}{\textbf{Class}}} & \multirow{2}[3]{*}{\textbf{Support}} & \multicolumn{5}{c||}{\textbf{Edit Distance Matching}} & \multicolumn{5}{c|}{\textbf{Exact Matching}} \\
			\cmidrule{3-12}          &       & \textbf{P} & \textbf{R} & \textbf{F1} & \textbf{AED} & \textbf{ANED} & \textbf{P} & \textbf{R} & \textbf{F1} & \textbf{AED} & \textbf{ANED} \\
			\midrule
			\midrule
			String & 207   & 0.957 & 0.953 & 0.955 & 1.795 & 0.120 & 1.000 & 0.775 & 0.802 & 4.144 & 0.223 \\
			Numbers & 60    & 0.971 & 0.971 & 0.971 & -     & -     & 0.971 & 0.971 & 0.971 & -     & - \\
			Algorithmic & 105   & 0.874 & 0.865 & 0.865 & 3.342 & 0.263 & 1.000 & 0.577 & 0.604 & 5.488 & 0.422 \\
			General & 76    & 0.702 & 0.530 & 0.567 & 6.926 & 0.365 & 0.974 & 0.419 & 0.507 & 7.564 & 0.506 \\
			\bottomrule
		\end{tabular}%
		\label{tab:perclass}%
	\end{table*}%
	
	\subsection{Class-Wise Performance and Error Analysis}

	In this section, we analyze the model's performance across each transformation class and highlight specific challenges encountered in some cases. This evaluation helps identify patterns in model performance and limitations that may inform future improvements. Table~\ref{tab:perclass} provides an overview of the results for each class, with support indicating the number of tables in each category. 
	
	In the string transformation class, the framework demonstrates strong performance. When edit-distance matching is applied, only a few cases are missed, primarily due to noisy data or substantial variations in formatting that were not adequately represented in the provided examples. Edit-distance matching allows for minor discrepancies, improving recall, though it occasionally results in missed rows when format variations exceed the threshold for an approximate match. 
	When exact matching is employed, the model's recall decreases in the string class. This drop is expected, as exact matching is more stringent and does not allow for minor discrepancies or transformations that partially transform the source into target. For instance, in a dataset containing information about state governors formatted as \texttt{Name; (birth date - death date)} in the source and \texttt{Name - (in office years)} in the target, the model successfully maps names but the date transformation is not feasible via string operations, which will adversely affect the performance when exact matching is employed.

	Moreover, in cases where provided examples are not sufficiently representative the model may generate a transformation function that fails to generalize. Consider top row in Figure~\ref{fig:etables} as an example, if the examples only include names without middle names, the model might generate a transformation that concatenates the first letter of the first name with the last name, failing to account for middle names when they appear. This indicates the limitations of random sample selector used in our  experiments and highlights the need for a more strategic approach to example selection, which is an area for future improvement.
	
	For numeric transformations, the model performs exceptionally well across all tables, achieving near-perfect precision and recall. This success can be attributed to the straightforward nature of numeric relationships, where mathematical functions can precisely capture the transformation patterns. The few cases where the model did not perform well were mostly outliers or instances where no meaningful relationship existed between the input and output values.
	In the algorithmic transformation class, while the overall performance remains strong, we observe a slight drop in both precision and recall. The primary challenge here is the complexity of certain algorithms. For example, transforming Gregorian dates to Hijri dates involves a sophisticated algorithm that the model partially captures. While it correctly identifies the relationship among source and target, the generated code may is incomplete, leading to some missed transformations.

	\begin{table*}[htbp]
		\centering
		\caption{Performance of the framework with Smaller and Open Source LLMs}
		\setlength\extrarowheight{1.5pt}
		\begin{tabular}{|c|c||c|c|c|c|c||c|c|c|c|c|}
			\toprule
			\multirow{2}[4]{*}{\textbf{Model}} & \multirow{2}[4]{*}{\textbf{Dataset}} & \multicolumn{5}{c||}{\textbf{Edit Distance Matching}} & \multicolumn{5}{c|}{\textbf{Exact Matching}} \\
			\cmidrule{3-12}          &       & \textbf{P} & \textbf{R} & \textbf{F1} & \textbf{AED} & \textbf{ANED} & \textbf{P} & \textbf{R} & \textbf{F1} & \textbf{AED} & \textbf{ANED} \\
			\midrule
			\midrule
			\multirow{4}[2]{*}{\begin{sideways}\textbf{GPT-4o}\end{sideways}} & Web Tables (WT) & 0.975 & 0.971 & \textbf{0.973} & 2.223 & 0.094 & 1.000 & 0.690 & \textbf{0.770} & 6.362 & 0.308 \\
			& SpreadSheet (SS) & 0.953 & 0.949 & 0.951 & 2.132 & 0.104 & 1.000 & 0.798 & \textbf{0.809} & 5.270 & 0.202 \\
			& Table Transformation (TT) & 0.919 & 0.918 & \textbf{0.918} & 2.466 & 0.219 & 0.992 & 0.733 & \textbf{0.756} & 3.915 & 0.344 \\
			& KBWT  & 0.722 & 0.545 & \textbf{0.580} & 7.139 & 0.387 & 0.975 & 0.445 & \textbf{0.523} & 7.674 & 0.516 \\
			\midrule
			\midrule
			\multirow{4}[2]{*}{\begin{sideways}\textbf{GPT-4o.m}\end{sideways}} & Web Tables (WT) & 0.924 & 0.920 & 0.922 & 3.752 & 0.142 & 0.983 & 0.626 & 0.685 & 8.214 & 0.372 \\
			& SpreadSheet (SS) & 0.979 & 0.976 & \textbf{0.977} & 1.763 & 0.072 & 1.000 & 0.802 & 0.806 & 4.730 & 0.198 \\
			& Table Transformation (TT) & 0.906 & 0.904 & 0.905 & 5.325 & 0.273 & 0.990 & 0.671 & 0.695 & 5.002 & 0.423 \\
			& KBWT  & 0.659 & 0.483 & 0.519 & 7.534 & 0.451 & 0.939 & 0.373 & 0.440 & 8.430 & 0.591 \\
			\midrule
			\midrule
			\multirow{4}[2]{*}{\begin{sideways}\textbf{LLaMA}\end{sideways}} & Web Tables (WT) & 0.822 & 0.820 & 0.821 & 6.837 & 0.275 & 0.994 & 0.437 & 0.491 & 11.298 & 0.568 \\
			& SpreadSheet (SS) & 0.915 & 0.912 & 0.913 & 3.551 & 0.235 & 0.981 & 0.389 & 0.398 & 8.942 & 0.611 \\
			& Table Transformation (TT) & 0.815 & 0.811 & 0.812 & 7.422 & 0.505 & 0.992 & 0.452 & 0.465 & 7.531 & 0.717 \\
			& KBWT  & 0.546 & 0.317 & 0.358 & 9.288 & 0.597 & 0.856 & 0.222 & 0.278 & 9.423 & 0.751 \\
			\bottomrule
		\end{tabular}%
		\label{tab:other_models}%
	\end{table*}%

	In the general transformation class, although the model significantly outperforms the baseline, the performance metrics are lower compared to the other classes. This transformation class is the most complex, as it heavily relies on the LLM's knowledge to produce accurate mappings, introducing several limitations. In some cases, the model has difficulty accurately identifying the types of the source and target columns. For instance, when the source is a U.S. Patent ID and the target is the patent name, the model may misinterpret the data, resulting in incorrect or nonsensical outputs.
	Even when the model correctly determines the column types, certain transformations in this class demand external knowledge beyond the model’s pre-trained data. When tasked with mapping SWIFT codes to bank names, the model correctly recognizes the column types but often produces inaccurate or hallucinated outputs, indicating a limitation in knowledge availability.
	
	Another common issue arises in one-to-many and many-to-many relationships, where multiple valid mappings exist for each source item. For example, in tables linking movies to their casts, a single movie may have numerous associated actors, and the model may generate a correct output that does not match the target, which is considered incorrect. These cases are not rare and contribute notably to the lower performance in the general class. Addressing these complex mappings effectively remains a challenge and is kept for future work.

	\subsection{Evaluation of Different Model Sizes and Open-Source Alternatives}
	
	To evaluate how the size and type of language models affect our framework's performance, we experimented with different commercial and open-source models of varying sizes. Table~\ref{tab:other_models} summarizes the results of these experiments. In this table, GPT-4o refers to the \texttt{gpt-4o-2024-05-13} model, one of the leading state-of-the-art commercial models. GPT-4o.m denotes the smaller and more cost-effective \texttt{gpt-4o-mini-2024-07-18} model. The LLaMA model represents the \texttt{Llama-3.1-8B-Instruct}, an open-source model with 8 billion parameters executable on a single GPU.

	As anticipated, the largest model, GPT-4o, generally outperforms the smaller models in terms of precision, recall, and F1-score, due to its advanced architecture and larger parameter count, especially when exact matching is used.
	However, for smaller models, edit-distance-based matching significantly mitigates performance gaps. This method accommodates minor variations in output, helping smaller models perform competitively. Notably, the gap between F1-scores using edit-distance versus exact matching grows as model size decreases. For instance, on the WT dataset, this difference is approximately 0.20 for GPT-4o, 0.24 for GPT-4o-mini, and 0.33 for LLaMA 3.1. The benefit of edit-distance matching even allows the GPT-4o-mini model to outperform GPT-4o on the SS dataset, underscoring its value in compensating for smaller models' limitations.

	Interestingly, the smallest model tested, LLaMA 3.1, demonstrates baseline-comparable or superior performance across datasets with edit-distance matching. Although there is a noticeable performance drop compared to the GPT models, the gap is not substantial in the Spreadsheet dataset. This outcome implies that for less complex tasks or in scenarios where computational resources are limited, smaller open-source models like LLaMA can be a viable alternative.
	However, on more complex datasets like KBWT, which require handling intricate transformations and may depend on external knowledge, the larger GPT-4o model maintains a clear advantage. Additionally, when exact matching is necessary---such as in applications requiring high accuracy without tolerance for errors---the larger models demonstrate superior performance.

	\section{Conclusion and Future works}
	In this paper, we introduced TabulaX, a novel framework that leverages LLMs for Multi-class Table Transformations. TabulaX addresses the challenges of integrating a wide range of heterogeneous tables with mismatched formats by classifying input data into distinct transformation classes---string-based, numerical, algorithmic, and general---and applying appropriate methods tailored to each class. Our approach generates human-interpretable transformation functions in the form of numeric formulas and programming code, enhancing transparency and allowing users to understand and modify the mappings as needed.
	Through extensive experiments on real-world datasets from various domains, we demonstrated that TabulaX outperforms existing state-of-the-art approaches in terms of accuracy and supports a broader class of transformations. %
	
	As possible future work, one direction is improving the handling of general transformations that require external knowledge; retrieval-based methods or domain-specific knowledge bases could be integrated to access additional information not present in the input data. Refinements to the classification mechanism could also be explored, potentially using more advanced models or leveraging metadata when available, to increase the accuracy of class detection. Additionally, extensions to support multi-column transformations, handling complex relational mappings, and further evaluations on downstream tasks such as anomaly detection, data imputation, and entity matching are among other potential future directions.

	\section*{Acknowledgments}
	This research was partially supported by the Natural Sciences and Engineering Research Council of Canada.
	
	\bibliographystyle{ACM-Reference-Format}
	\bibliography{references}

\end{document}